\documentclass[a4paper,12pt,english,german]{article}

\usepackage{bbm}
\usepackage{amsmath}
\usepackage{amsmath,amssymb}
\usepackage{graphicx}

\begin{document}

\centerline{ \bf A local-time-induced unique pointer basis}

\bigskip

\centerline{
J. Jekni\' c-Dugi\' c$^{1}$, M. Arsenijevi\' c$^{2}$ and M. Dugi\'
c$^{2}$}

\bigskip

\centerline{$^{1}$Department of Physics, Faculty of Sciences and
Mathematics, 18000 Ni\v s, Serbia}

\centerline{ $^{2}$Department of Physics, Faculty of Science,
34000 Kragujevac, Serbia}

\bigskip

{\bf Abstract} There is a solution to the problem of asymptotic
completeness in many body scattering theory that offers a specific
view of the quantum unitary dynamics which allows for the
straightforward introduction of local time for every, at least
approxi\-ma\-te\-ly closed, many-particle system.  In this
approach, Time appears as a hidden classical parameter of the
unitary dynamics of a many-particle system. We show
 that  a closed many-particle system can
exhibit behavior that is characteristic for open quantum systems
and there is no need for the "state collapse" or environmental
influence. On the other hand,  closed few-particle systems bear
high quantum coherence.  This local time scheme encompasses
concepts including "emergent time", "relational time" as well as
the "hybrid system" models with possibly induced gravitational
uncertainty of time.

\bigskip

{\bf 1. Introduction}

\bigskip

A solution to the  problem of asymptotic completeness in the
many-body scattering theory offers a specific view of the
 quantum unitary dynamics. The important work of Enss [1,2] opened
the door for new methods in solving the problem. On this basis,
the later elaboration due to Kitada [3,4] allowed Kitada [5-7] to
introduce the notion of local time, that is a dynamics generated
by the Hamiltonian of the local system that can serve as a [local]
"clock".

The notion of local time or "multi-time" is not a new idea. Mainly
motivated by relativity, a separate time coordinate for every
particle in a composite system has been introduced, see e.g.
Refs.[8,9] and the references therein. It is also shown that the
"timeless" Wheeler-DeWitt equation:

\begin{equation}
H(x) \vert \Psi\rangle = 0,
\end{equation}

\noindent follows if one assumes the existence of a preferred
foliating family of spacelike surfaces in spacetime [10].

On closer inspection, such {\it ad hoc} schemes for local time
face serious obstacles in nonrelativistic quantum context. For
example, it can be shown, that "Multi-time Schr\" odinger
equations cannot contain interaction potentials" [8].
Consequently, the following scenario appears to be inescapable:
non-relativistic Universe, seen as a collection of interacting
subsystems, must have unique, global time that is common for all
possible subsystems.  In other words, it would seem that there is
no room for the idea of local time in non-relativistic quantum
theory; regarding the relativistic considerations see Ref.[11].

However, Kitada's [5] concept of local time is neither {\it ad
hoc} nor does it suffer from such problems. Based on the
non-relativistic many-body scattering theory, Kitada's approach
defines ''local systems (clocks)" as approximately isolated from
the other systems in the sense that a local clock of each local
system is described by the Schr\" odinger law. {\it Prima facie},
the assumption on independent local clocks may seem to be a
reminiscent of the observation [8] that interactions (and the
induced quantum entanglement) may ruin the idea of local time in
non-relativistic quantum theory. However, in Kitada's scheme, a
particle's time is {\it shared with some other particles} and is
{\it not fixed} once for all. The particle's time is the time
defined by the isolated local system that the particle is a part
of. Dynamics of the total system (the Universe) can place the
particle in different local systems and therefore in different
local times. This provides a specific interpretation of eq.(1) --
on the level of the total universe, time does not exist, but on
the local level, time does exist [3-7] -- as well as makes the
scheme well suited both for the purposes of quantum measurement
[12-14] and decoherence [14-18] and for some models of open
quantum systems [19,20] theory. A clear example of the latter
which we know from measurement and decoherence is that the
conglomeration of small few-particle systems can produce, at some
point, a many-particle system that is approximately isolated from
the other ``conglomerates".

In quantum measurement, one deals with the many-particle systems
that are assumed to be almost isolated (closed) systems. That is,
an "object of measurement + apparatus" ($O+A$) system, or "object
of measurement + apparatus + the apparatus' environment"
($O+A+E$), is subjected to unitary Schr\" odinger dynamics
disregarding existence of other such systems. In the standard
decoherence theory, the composite system "open system +
environment" ($S+E$) is assumed to be subject to  unitary Schr\"
odinger dynamics despite the presence of other open systems and
their environments. Finally, in the context of the open quantum
systems theory, it is known that virtually every dynamics of an
open system $S$ can be described by the Schr\" odinger law for the
extended system $S+E$ [20]. So, the macroscopic (many-particle)
systems $O+A$, $O+A+E$ and $S+E$ {\it that are subject to the
Schr\" odinger law} appear as  perfect candidates for application
of, and some kind of a test for, the local time scheme of Kitada.

With this motivation in mind, we hypothesize the following
 rule for the universally valid quantum theory:
 "Every many-particle system that is [approximately] subject to the Schr\" odinger law, can
be assigned  a local time independently of other such systems",
and we investigate the consequences for the description of the
quantum-decoherence-like processes.

In this paper we slightly extend the original proposal [5] by
pointing out uncertainty of local  time for the local clocks. This
introduces the local time  effectively as a hidden classical
parameter in the system's unitary Schr\" odinger dynamics. We
introduce a Gaussian distribution for a time interval as the
obvious choice and for a subsystem of a closed bipartite
many-particle system we obtain {\it unique} "pointer basis". Our
results come from the {\it macroscopic} domain but without a need
to resorting to "state collapse" or environmental influence.

Implications of  the local-time scheme of Kitada  are noteworthy.
Hence we believe it is worth further investigation in the
foundations of quantum theory as well as towards the original
relativistic motivations with the view of the possible reduction
of the gap between the quantum and relativistic theories.

The contents of this paper are as follows. In Section 2 we provide
a brief account of the many-particle scattering theory and
recapitulate Kitada's notion of local time. In Section 3 we point
out uncertainty of time for bipartition of a many-particle system
and set the quantitative criteria  that revolve around eq.(9). In
Section 4 we apply Kitada's scheme to the standard
decoherence-like processes; Section 4.2 provides the main result
of our paper. Section 5 provides some illustrative examples that
exhibit technical simplicity, generality and clarity of the
local-time scheme in the context of quantum measurement and
decoherence. Section 6 is discussion of interpretational
ramifications of the local-time scheme and Section 7 is
conclusion.

\bigskip

{\bf 2. Outlook of the many-body scattering theory and the notion
of local time}

\bigskip

{\bf 2.1 Asymptotic completeness in the many-body scattering
theory}

\bigskip

A reader uninterested in details of the many-body scattering can
skip to Section 2.2. Scattering theory is essentially
time-independent perturbation theory applied to the case of a
continuous spectrum. The goal of scattering theory is to solve the
full energy-eigenstate problem

\begin{equation}
(E - H_{\circ} - V) \vert \Psi\rangle = 0,
\end{equation}

\noindent where $E > 0$ (unless otherwise specified), and $\vert
\Psi\rangle$ is the eigenstate of the full Hamiltonian $H = H_0 +
V$ with energy $E$. Already two-particles scattering is a hard
problem. The many-body scattering poses even the more serious
technical problems. It is due to Enss [1,2] that the method of
clustering the composite system in conjunction with the so-called
micro-local analysis method offers a systematic approach to the
problem for both short-range and long-range interactions (denoted
by $V$ in eq.(2)). Subsequent development of the field of
many-body scattering can be found e.g. in Sigal [21] and the
references therein.

The method of clustering consists in the following idea. A
composite, closed system $\mathcal{S}$ that consists of $N$
nonidentical particles, can be differently structured [22], e.g.
clustered in mutually non-intersecting clusters. E.g. a tripartite
system $\mathcal{S} = 1+2+3$ can be structured as: $\mathcal{S}_1
= \{1,2,3\}, \mathcal{S}_2 = \{ \{1\}, \{2,3\}  \}, \mathcal{S}_3
= \{   \{2\}, \{1,3\} \}, \mathcal{S}_4 = \{\{3\}, \{1,2\}\}$,
$\mathcal{S}_5 = \{\{1\}, \{2\}, \{3\}\}$ where the brackets "$\{
\ast \}$" denote one cluster. So, the structures $\mathcal{S}_i, i
= 2,3,4$ are different bipartitions of the total system
$\mathcal{S}$, the $\mathcal{S}_5$ is a tripartite structure of
the total system $\mathcal{S}$ while the $\mathcal{S}_1$
represents a formally unstructured system (only one cluster). For
every cluster, a center-of-mass ($CM$) and the relative-positions
($R$) degrees of freedom are introduced; the $R$ system's degrees
of freedom are often chosen as the Jacobi relative coordinates.
Then, bearing in mind the variety of the different possible
structures,
 {\it all} the possible scattering scenarios are described by the
scattering of the clusters' $CM$-systems.

For the $b$th structure (cluster decomposition) with $k$ clusters,
the collective relative positions variable $x_b = \{x_{bi},
i=1,2,3,...,k\}$ and the related conjugate momentum, $p_b =
\{p_{bi}, i=1,2,3,...,k\}$, are introduced for the clusters'
centers-of-mass systems. The commutators $[x_{bi}, p_{b'j}] =
\imath \hbar \delta_{ij} \delta_{bb'}$. The rest of the
relative-positions variables are collectively denoted by $x^b$
with the conjugate momentums $p^b$. Then the Hilbert state space
factorizes:

\begin{equation}
\mathcal{H} = \mathcal{H}_{CM} \otimes \mathcal{H}_b \otimes
\mathcal{H}^b.
\end{equation}

Of course, the factorization eq.(3) is different for  different
structures, i.e. $\mathcal{H}_b \neq \mathcal{H}_{b'}$. By placing
the reference frame in the total-system's $CM$ system, which is
common for all structures, i.e. by choosing $X_{CM} = 0$, the
factorization eq.(3) is reduced:

\begin{equation}
\mathcal{H} = \mathcal{H}_b \otimes \mathcal{H}^b.
\end{equation}

 Therefore, observation of the
scattering process reduces to observation  of the inter-cluster
Jacobi coordinates $x_{bi}$, for every structure $b$.

Another essential point in Refs.[1,2] comes from the fact that, as
emphasized above, scattering refers to the continuous spectrum of
the total system's Hamiltonian. So, the pure point-spectrum of the
Hamiltonian  should be removed from the consideration. The Enss'
solution is remarkable: he considers the "velocity operator" $v_b
= m_b^{-1}p_b$, where $p_b$ is defined above and $m_b$ is the
diagonal mass matrix with the diagonals being the intercluster
reduced masses, and introduces a kind of projection operator
$\tilde{P}_b^{M^m_b}$, that can be found in Appendix. Then for
every quantum state
 $\vert \Psi\rangle$ and for a wide class of the  potentials $V$, eq.(2),  Enss
proved:

\begin{equation}
\left\Vert \left({x_b \over t_m} - v_b\right) \tilde{P}_b^{M^m_b}
\exp(-\imath t_m H/\hbar) \vert \Psi\rangle \right\Vert \to 0,
\end{equation}

\noindent as the time-index $m \to \pm\infty$, for {\it every}
structure $b$ save for the "structureless" one with only one
cluster ($k$ = 1). The norm $\Vert \vert \psi\rangle\Vert =
\sqrt{\langle \psi\vert \psi\rangle}$, where $\langle \psi\vert
\psi\rangle$ is the scalar product in the Hilbert state space
$\mathcal{H}$, and the index $m$ is an integer taking positive or
negative values including zero. The limit in eq.(5) is in the
standard sense of limit for a sequence of real norms indexed by
$m$.

{\bf 2.2 The notion of local time}

\bigskip

Physical meaning of eq.(5) is as follows. For a single free
particle:

\begin{equation}
(x - t v) \exp(-\imath t T / \hbar) = \exp(-\imath t T / \hbar) x,
\end{equation}

\noindent where $x$ is the position and $v$ the velocity
observable, while $T$ stands for the kinetic-energy observable and
$t$ is an instant of time. Action of the operators in eq.(6) on a
wave function $\Psi(x)$ is as follows: If the support of $\Psi(x)$
is around some $x_{\circ}$ in the instant $t_{\circ} = 0$, then
the support of the propagated wave function $\exp(-\imath t T /
\hbar ) \Psi(x)$ is localized around the point $x_{\circ} + v t$
in the instant of time $t$. Eq.(5) essentially says that {\it the
same holds} for the wave function of a {\it closed many-body}
(many-particle) system with virtually arbitrary kinds of pair
interactions in the system.

Eq.(5) encompasses all the possible scattering scenarios for the
closed system $\mathcal{S}$ as both the Hamiltonian, $H$, and the
time instants, $t_m$, are {\it common for all structures}.
Therefore, measurement of arbitrary $x_b$ and $v_b$ and obtaining
their mean values, $\langle x_b\rangle$ and $\langle v_b\rangle$,
provides, at least sketchily, the measurement of time for the
total system [5-7]:

\begin{equation}
{\langle x_b\rangle \over \langle v_b\rangle} \sim t,
\end{equation}

\noindent in the asymptotic limit. This allows for the definition
of the notion of $t$ as "the reading of a clock $b$" in a way
consistent with informal discussions of Relativity.

Scattering is a fundamental method of interaction for systems at
all quantum scales. In analogy with eq.(6), it is therefore
reasonable to interpret eq.(7) as a {\it notion} of time, which is
common for all the structures, clusters and particles in the {\it
closed} system $\mathcal{S}$, but {\it not necessarily} for some
other {\it closed} systems, $\mathcal{S'}$, $\mathcal{S}''$ etc.
The different rates of operation give rise to the intuitive
picture of time as a characteristic of a local, i.e. approximately
closed system.

While the concept of local time in certain schemes is an {\it ad
hoc} idea  [8,9], in the many-body scattering theory, this notion
naturally fits with eq.(5). Eq.(5) directly provides the following
rules  [5]: (a) Systems with different Hamiltonians such as those
with a different numbers of particles, or different kinds of
particles, or different kinds of interactions between the
particles are subject to different local times; (b) Systems that
mutually interact are subjected to the same time; (c)
Noninteracting systems need not have a common time; (d)
Nonidentical many-body systems which do not interact and locally
follow independent Schr\" odinger dynamics do not have a common
time--which makes the universal time undefinable, as for eq.(1);
(e)  Local time refer even to the mutually identical many-body
systems, as long as they represent the mutually independent local
systems.

In the remainder of this paper, we use the above points (a)-(e) as
a matter of principle i.e. as the new {\it universal rule} in
quantum theory.

{\bf 3. Uncertainty of local time}

\bigskip

In certain processes, such as atomic collisions and chemical
reactions, there may occur a change in the system's structure,
$\mathcal{S}_b \to \mathcal{S}_{b'}$ [22,23].  In the case of more
fundamental nonrelativistic particles scattering experiments, a
structure $\mathcal{S}_b$ typically remains unchanged. Then the
measurement of the intercluster observable $x_b$ describes
collisions of the particles for that structure.

Within the standard universally valid quantum mechanics, a closed
system is defined by the unique state [in the Schr\" odinger
picture]:

\begin{equation}
\vert \Psi (t_{\circ})\rangle = U(t_{\circ}) \vert
\Psi(t=0)\rangle,
\end{equation}

\noindent where  $U(t) = \exp(-\imath t H/\hbar)$ and $H$ is the
total system's Hamiltonian. Of course, eq.(8) assumes unique,
global physical time. If eq.(8) models a measurement (or
decoherence), then the measurement is assumed to be complete in an
instant $t_{\circ}$, and the limit $t_{\circ} \to \infty$ is
formally allowed.

According to the  point (e), Section 2.2, even in the limit of
zero metrological error, {\it there is a time uncertainty} $\Delta
t$ in determining the {\it finite}  $t_{\circ}$ that gives instead
of eq.(8):

\begin{equation}
\sigma = \int_{t_{\circ} - \Delta t}^{t_{\circ} + \Delta t}
\rho(t) \vert \Psi(t)\rangle\langle \Psi(t) \vert dt,
\end{equation}

\noindent where for the time probability-density $\rho(t)$:

\begin{equation}
\int_{t_{\circ} - \Delta t}^{t_{\circ} + \Delta t} \rho(t) dt =1,
\int_{t_{\circ} - \Delta t}^{t_{\circ} + \Delta t} t \rho(t) dt =
t_{\circ}.
\end{equation}

For the time probability-density we require: (1) to be symmetric
on the narrow interval $[t_{\circ} - \Delta t, t_{\circ} + \Delta
t]$, (2) regarding the decoherence-like processes, see eq.(20)
below,  $t_0$ is the time instant in which decoherence
 is approximately complete, and therefore the limit $t_{\circ} \to
\infty$ should be formally allowed, and (3) to allow a proper
limit $\rho(t) \to \delta(t)$, with the Dirac delta function
$\delta (t)$, in order to be reducible to the standard case
eq.(8).

Physically, and also operationally, i.e. for an observer, the
state eq.(9) is {\it objective}--the so-called "proper mixture".
Determining the time instant $t$ from the interval $[t_{\circ} -
\Delta t, t_{\circ} + \Delta t]$ is equivalent with distinguishing
between non-orthogonal states $\vert \Psi(t)\rangle$. Hence the
no-cloning theorem [24] makes the task of distinguishing the time
instants from the interval $[t_{\circ}-\Delta t, t_{\circ} +
\Delta t]$ impossible {\it in principle} [25].

The time uncertainty $\Delta t$ does not introduce uncertainty of
energy. Every term in eq.(9) describes a unitary Schr\" odinger
evolution with energy preservation: $\langle \Psi(t) \vert H \vert
\Psi(t)\rangle = \langle \Psi(t=0) \vert H \vert
\Psi(t=0)\rangle$. Then there is energy conservation also for the
state eq.(9): $tr \sigma H = const$.

\bigskip

{\bf 3.1 The state eq.(9) is mixed}

\bigskip

By construction, the state eq.(9) is mixed. Nevertheless, for the
arbitrarily short interval $\Delta t \ll t_{\circ}$, from
eqs.(10),(11):

\begin{equation}
\sigma \approx \int_{t_{\circ} - \Delta t}^{t_{\circ} + \Delta t}
\rho(t) \left(I - {\imath (t - t_{\circ}) \over \hbar} H \right)
\vert \Psi(t_{\circ})\rangle\langle \Psi(t_{\circ}) \vert \left(I
+ {\imath (t - t_{\circ}) \over \hbar}  H\right) dt \approx\vert
\Psi(t_{\circ}) \rangle\langle \Psi(t_{\circ}) \vert
\end{equation}

\noindent with an error of the order of $(\delta t/\hbar)^2$ and
the standard deviation $\delta t$. For $t_{\circ} \gg 1$ [cf. the
above point (2)], the interval $\Delta t$ need not be that short
while it can still fulfill $\Delta t \ll t_{\circ}$.

On the other hand, for $\Delta t
> \tau_{min} = \max\{\pi\hbar/2\Delta H, \pi \hbar / 2(\langle H\rangle_{t=0} - E_{g})\}$, where $\Delta H$
is the standard deviation and  $E_g $ stands for the Hamiltonian
ground energy,
 there are  three time instants, $t_{\circ} - \Delta t$, $t_{\circ}$
and $t_{\circ} + \Delta t$, which pertain to mutually
[approximately] orthogonal states [26,27]. While $\Delta t$ can be
very small in some physical units, it can still be "large" so as,
for the coarse grained time axis with the width $\Delta t$, the
state eq.(9) reads:

\begin{equation}
\sigma = p_{-} \vert \Psi(t_{\circ} - \Delta t)\rangle\langle
\Psi(t_{\circ} - \Delta t)\vert + p_{\circ} \vert
\Psi(t_{\circ})\rangle\langle \Psi(t_{\circ} )\vert + p_{+} \vert
\Psi(t_{\circ} + \Delta t)\rangle\langle \Psi(t_{\circ} + \Delta
t)\vert.
\end{equation}

For such time interval $\Delta t$, the states in eq.(12) can be
mutually distinguished.

In this paper we reduce our attention to the proper small
intervals $\Delta t$ that allow the limit $\Delta t \to 0$ i.e.
only slight deviation from eq.(8), while not leading either to
eq.(11) or to eq.(12).

In accordance with eqs.(10)-(11) and due to the above points
(1)-(3), we choose a  Gaussian time probability-distribution:

\begin{equation}
\rho(t) = {\sqrt{\lambda \over \pi}} \exp(-\lambda (t -
t_{\circ})^2),
\end{equation}

\noindent which in the limit $\lambda \to \infty$ provides the
standard case eq.(8). Therefore, we choose the smallest possible
$\lambda$ so as $\tau_{min}/2
> \Delta t > \lambda^{-1}$ and:

\begin{equation}
\int_{t_{\circ} - \Delta t}^{t_{\circ} + \Delta t} \rho(t) dt
\approx \int_{-\infty}^{\infty} \rho(t) dt =1.
\end{equation}

\noindent Eqs.(13) and (14) provide the estimate $\Delta t
\succsim \lambda^{-1/2}$ that implies $\Delta t > \lambda^{-1}$
and the above constraint reduces to $\tau_{min} > 2\Delta t$.

The choice of a Gaussian density eq.(13) is by no means the only
one possible. Nevertheless, it facilitates the analysis and allows
a comparison of the different models of the composite $O+A$
system--see Section 5 for some relevant models. The possible
extensions of our considerations toward non-Gaussian or even
model-dependent density $\rho(t)$ may be physically sound but are
beyond the scope of the present paper.

We are interested in the description of quantum decoherence which
includes finite-dimensional systems. In this context, typically,
the pure discrete (point) energy spectrum and bound states are
considered: the exact spectral form $H = \sum_n h_n \vert
n\rangle\langle n \vert$ describes a closed system confined to a
finite region of space and involving a finite number of particles.

 Then for arbitrary initial pure state $\vert
\Phi \rangle = \sum_n c_n \vert n \rangle$ eq.(8) reads:

\begin{equation}
U(t_{\circ}) \sum_n c_n \vert n \rangle = \sum_n c_n \exp(-\imath
t_{\circ} h_n/\hbar) \vert n\rangle.
\end{equation}

Now eq.(9) takes the form:

\begin{equation}
\sigma =  \sum_n \vert c_n\vert^2 \vert n\rangle\langle n\vert +
  \sum_{n\neq
n'} c_n c_{n'}^{\ast}  \exp(-\imath t_{\circ}(h_n - h_{n'})/\hbar)
\exp(-(h_n - h_{n'})^2/4\hbar^2\lambda) \vert n\rangle\langle
n'\vert.
\end{equation}

\noindent In calculating eq.(16) we used the  Gaussian integral:
$\int_{-\infty}^{\infty} \exp(-ax^2/2 + \imath Jx) dx =
(2\pi/a)^{1/2}\times$ $\exp(-J^2/2a)$, where $a > 0$ and $J$ are
real numbers with $J$ being conjugate variable of $x$.  By very
definition eq.(9), the state $\sigma$ is hermitean, positive and
with unit trace.

From eq.(16):

\begin{equation}
tr \sigma^2 = \sum_{n,n'} \vert c_n\vert^2 \vert c_{n'}\vert^2
\exp (-(h_n - h_{n'})/2\hbar^2\lambda) < 1,
\end{equation}

\noindent that clearly exhibits: the state $\sigma$ is mixed.

\bigskip

{\bf 3.2 Few-particle versus many-particle systems}

\bigskip

The terms $\exp(-(h_n - h_{n'})^2/4\hbar^2\lambda)$ appearing in
eq.(16) can in general vary from almost $0$ to $1$. There can be
plenty of close energy values  and thus plenty of terms in the sum
eq.(16) can equal or be very close to 1. For poor energy spectrum,
which is characteristic for some small (few-particle) systems,
eq.(17) can be very close to 1, i.e. there can be pure states in
close vicinity of the mixed $\sigma$ state. This is readily seen
for the standard state eq.(8) and eq.(15): the fidelity [25],
$\mathcal{F} = tr\sqrt{\vert \Psi\rangle\langle \Psi\vert \sigma
\vert \Psi\rangle\langle \Psi\vert} = \sqrt{\langle \Psi \vert
\sigma \vert \Psi\rangle } = \sqrt{\sum_{n,n'} \vert c_n\vert^2
\vert c_{n'}\vert^2 \exp (-(h_n - h_{n'})/4\hbar^2\lambda)}$.

Therefore, high quantum coherence in the total $O+A$ system can be
expected as a consequence of the constraint $\tau_{min}/2
> \Delta t$ and eq.(14). To this end appear the
following two questions. First, whether mixed state eq.(9)
regarding few-particle systems, such as e.g. the EPR pairs, could
be in conflict with phenomenology? And, the second, whether one
can safely use pure states in the vicinity of the mixed state
$\sigma$ for many-particle  systems?

The first  question appears in the context of the decoherence
theory and the open system theory: how can we reproduce validity
of the Schr\" odinger law, i.e. quantum coherence, on the
microscopic level?  The often offered answer is pragmatic: the
small systems are very well isolated and the environmental
influence is {\it almost} negligible {\it in practice}, see e.g.
[13]. Hence the local-time scheme goes along with the standard
decoherence theory in describing the few-particle systems [13,14]:
the mixing of states   can be weak and the  small system can be
considered to be in pure state for the most of the practical
purposes.

On the other hand, bearing in mind  that the energy spectrum for
many-particle systems is dense, the
 typical macroscopic measurements  bury the exact
 eigenvalues and provide a seemingly
continuous spectrum--the very basis of the "continuous
approximation" that is  widely used e.g. in condensed matter
physics. Then eq.(9) reads:

\begin{equation}
\sigma = \int dE dE' \Psi(E) \Psi^{\ast}(E') \exp(-\imath
t_{\circ} (E - E')/\hbar) \exp[-(E - E')^2/4\hbar^2\lambda] \vert
E \rangle \langle E' \vert.
\end{equation}

Eq.(18) resembles the well known expressions for the
continuous-variable (CV) systems decoherence: While there is high
coherence over all in the state $\sigma$, eq.(18), there is
substantial loss of coherence for certain energy values, for which
$\exp(-(E - E')^2/4\hbar^2\lambda) \ll 1$. This situation is
typical for virtually all CV open systems [13,14,16,18].  Bearing
in mind that this cannot {\it in principle} be achieved by pure
states, we obtain the answer to the above-posed second question:
no, the use of the "close" pure states in general cannot be
useful.

Coarse graining of the energy spectrum can reduce coherence in the
closed system. By definition, coarse graining decreases the number
of energy eigenvalues as well as the number of Gaussian terms,
$\exp(-(h_n - h_{n'})^2/4\hbar^2 \lambda)$, with smaller terms
$\vert h_n - h_{n'}\vert$ in the exponent. For certain
few-particle systems with poor spectrum, typically, this procedure
will not work. Therefore while  the few-particle systems can be
expected to exhibit approximate quantum behavior,  the
many-particle systems can exhibit quantal versus classical-like
behavior--the latter being a reminiscent of the conjecture that
this is not merely a matter of the system's spatial size or mass
but rather of the energy scale [28]. Some examples can be found in
Section 5.

\bigskip

{\bf 3.3 Local-time as a dynamical map}

\bigskip

It cannot be overemphasized that the mixed state eq.(9) refers to
a {\it closed} system. Compared to eq.(8), eq.(9) emphasizes the
following  map:

\begin{equation}
\nonumber \vert \Psi(t=0)\rangle\langle \Psi(t=0) \vert \to \sigma
(t_{\circ}) = \int_{t_{\circ} - \Delta t}^{t_{\circ} + \Delta t}
\rho(t) \vert \Psi(t)\rangle\langle \Psi(t) \vert dt.
\end{equation}

It is readily extendible to the following dynamical map:

\begin{equation}
\nonumber \mathcal{S}(t = 0) = \sum_i p_i
\vert\Psi_i(t=0)\rangle\langle \Psi_i(t=0) \vert \to
\mathcal{S}(t_{\circ}) = \sum_i p_i \sigma_i (t_{\circ})
\end{equation}

\noindent where every $\sigma_i (t_{\circ})$ is of the form of
eq.(9). Since the $\sigma$, eq.(9), is Hermitean, positive and
with unit trace, this dynamical map is positive. This map is an
instance of the "random unitary evolution" that is known for the
finite-dimensional systems to be completely positive [29].

\bigskip

{\bf 4. A local-time scheme for decoherence-like processes}

\bigskip

In this section, our analysis is a deductive application of the
rules established in Sections 2.2 and 3 for the case of strong
interaction in the $O+A$ system. We also briefly analyze quantum
measurement and reproduce some well-known results.

\bigskip

{\bf 4.1 Strong interaction in the $O+A$ system}

\bigskip

The points (a)-(e) in Section 2.2 set the clear-cut scenario of
the decoherence-like situations. Before interaction, the  systems
$O$ and $A$ are described by the Hamiltonian $H_O + H_A$.
According to the point (b), the systems may or may be not
subjected to the same time.
 However, interaction in the $O+A$ system introduces the new Hamiltonian, $H_O+H_A+H_{int}$,  where
$H_{int}$ is the interaction Hamiltonian. According to the point
(b), now both the $O$ and $A$ systems are subject to the same
local time. According to the point (a), the time for the $O+A$
system is not the same as for the $O$ i.e. the $A$ system
 before interaction. So, the start of the interaction locally defines the
 initial time instant, $t$, for the {\it newly formed many-body}
 $O+A$ {\it system} and sets the "clock" implemented by this system to the
 value $t = 0$. As long as the system $O$ and the apparatus $A$ are precisely defined, the time instant $t=0$ is assumed
 to be uniquely defined. In the terms of the standard theory, the local $t=0$ corresponds to an instant $t_{\circ}$
 that is assumed to be arbitrary but fixed and measured by a clock at the observer's disposal.

 In quantum decoherence (and also in measurement), typically,  interaction  in the
$O+A$ system is assumed to dominate the system's dynamics [13-18].
Physically it means that the self-Hamiltonian can be neglected: $H
= H_O + H_A + H_{int} \approx H_{int}$.

We  consider a pure initial tensor-product state $\vert
\phi\rangle_O \vert \chi\rangle_A$. The separable spectral form
for the interaction Hamiltonian with the real eginevalues
$h_{\alpha\beta}$ [13-18,30]:

\begin{equation}
H_{int} = \sum_{\alpha,\beta} h_{\alpha\beta} P^O_{\alpha} \otimes
\Pi^A_{\beta},
\end{equation}

\noindent where appear the projectors, $P$ and $\Pi$, on the
respective factor spaces.

Then the standard unitary unitary dynamics gives the pure state:

\begin{equation}
\vert \Psi(t)\rangle = \sum_{\alpha} b_{\alpha} \vert
\alpha\rangle_O \vert \chi_{\alpha}(t)\rangle_A,
\end{equation}

\noindent where

\begin{equation}
\vert \chi_\alpha(t)\rangle_A = \sum_{\beta} d_{\beta}
\exp(-\imath t h_{\alpha\beta}/\hbar) \vert \beta\rangle_A;
\end{equation}

\noindent with $b_{\alpha} \vert \alpha\rangle_O = P^O_{\alpha}
\vert \phi\rangle_O$ and $d_{\beta} \vert \beta\rangle_A =
\Pi^A_{\beta} \vert \chi\rangle_A$; $\sum_{\alpha} \vert
b_{\alpha}\vert^2 = 1 = \sum_{\beta} \vert d_{\beta} \vert^2$.

Substituting eq.(21) into eq.(9):

\begin{equation}
\sigma =  \sum_{\alpha} \vert b_{\alpha} \vert^2 \vert
\alpha\rangle_O\langle \alpha \vert \otimes \rho^A_{\alpha}
(t_{\circ}) + \sum_{\alpha\neq\alpha'} b_{\alpha}
b_{\alpha'}^{\ast} \vert \alpha\rangle_O\langle \alpha' \vert
\otimes \rho^A_{\alpha\alpha'}(t_{\circ}),
\end{equation}

\noindent where $t_{\circ}$ is the time instant for which
decoherence is (at least approximately) complete.

In eq.(23):

\begin{equation}
\rho^A_{\alpha} = \sum_{\beta,\beta'} d_\beta d_{\beta'}^{\ast}
\exp(-\imath t_{\circ}(h_{\alpha \beta} - h_{\alpha
\beta'})/\hbar) \exp(-(h_{\alpha \beta} - h_{\alpha
\beta'})^2/4\hbar^2\lambda) \vert \beta\rangle_A\langle
\beta'\vert,
\end{equation}
\begin{equation}
\rho^A_{\alpha\alpha'} = \sum_{\beta,\beta'} d_\beta
d_{\beta'}^{\ast} \exp(-\imath t_{\circ}(h_{\alpha \beta} -
h_{\alpha'\beta'})/\hbar) \exp(-(h_{\alpha \beta} -
h_{\alpha'\beta'})^2/4\hbar^2\lambda) \vert \beta\rangle_A\langle
\beta'\vert.
\end{equation}

It is easy to see, that $\rho^A_{\alpha}$s are hermitean and
positive with unit trace.

\noindent {\bf Lemma 4.1} (i) {\it The density matrices}
$\rho^A_{\alpha}$ {\it are mutually approximately orthogonal for
most of the large values of} $t_{\circ}$,  {\it symbolically}
$\lim_{t_{\circ} \to \infty}\rho^A_{\alpha} \rho^A_{\alpha'}
\approx 0, \forall{\alpha\neq \alpha'}$; (ii) {\it the trace of}
$\rho^A_{\alpha\alpha'}$s {\it equals approximately zero for most
of the large values of} $t_{\circ}$, {\it  symbolically}
$\lim_{t_{\circ} \to \infty} tr_A \rho^A_{\alpha\alpha'} \approx
0, \forall{\alpha\neq \alpha'}$.

\noindent {\it Proof.} (i) From eq.(24), the matrix elements: 
\begin{eqnarray}
&\nonumber&  (\rho^A_{\alpha} \rho^A_{\alpha'})_{\beta\beta''} =
d_\beta d_{\beta''}^{\ast} \exp(-\imath t_{\circ}(h_{\alpha \beta}
- h_{\alpha'\beta''})/\hbar) \sum_{\beta'} \vert d_{\beta'}\vert^2
\exp(-\imath t_{\circ}(h_{\alpha' \beta'} -h_{\alpha
\beta'})/\hbar) \times\\&& \exp\{-[(h_{\alpha \beta} - h_{\alpha
\beta'})^2 + (h_{\alpha'\beta'} -
h_{\alpha'\beta''})^2]/4\hbar^2\lambda\}\nonumber\\&& \equiv
d_\beta d_{\beta''}^{\ast} \exp(-\imath t_{\circ}(h_{\alpha \beta}
- h_{\alpha'\beta''})/\hbar) \zeta \chi.
\end{eqnarray}

In the last row of eq.(26) we simplify notation and introduce: $0
< \epsilon_{\beta'} \equiv \exp\{-[(h_{\alpha \beta} - h_{\alpha
\beta'})^2 + (h_{\alpha'\beta'} -
h_{\alpha'\beta''})^2]/4\hbar^2\lambda\} \le 1$, $\zeta \equiv
\sum_{\beta'} \vert d_{\beta'}\vert^2 \epsilon_{\beta'}$,
$p_{\beta'} \equiv \vert d_{\beta'}
\vert^2\epsilon_{\beta'}/\zeta$ and $\omega_{\beta'} \equiv
(h_{\alpha \beta'} - h_{\alpha'\beta'})/\hbar$. Since
$\sum_{\beta'} p_{\beta'} = 1$, $\chi \equiv \sum_{\beta'}
p_{\beta'} \exp(-\imath t_{\circ} \omega_{\beta'})$ is the
well-known "correlation amplitude" [15]. For sufficiently long
time interval $[t, t+T]$, such that $t_{\circ} \in [t, t+T]$, for
$\alpha\neq\alpha'$, the correlation amplitude satisfies [15]: (a)
the time average on the interval $\lim_{T\to\infty} \langle
\chi\rangle_T = 0$, and (b) the standard deviation on the interval
$\lim_{T\to\infty} \langle \vert \chi \vert^2\rangle_T = 0$ for
typical models of the many-particle $A$ system.
 Bearing in mind that $\zeta \le 1$, the
point (i) is proved.

(ii) From eq.(25):

\begin{equation}
 tr_A \rho^A_{\alpha\alpha'} =
\sum_\beta \vert d_\beta\vert^2 \exp(-\imath t_{\circ}(h_{\alpha
\beta} - h_{\alpha'\beta})/\hbar) \exp(-(h_{\alpha \beta} -
h_{\alpha'\beta})^2/4\hbar^2\lambda),
\end{equation}

\noindent where the right hand side is of the $\chi$-function form
 considered in (i) above--that completes the proof
of the point (ii). \hfill Q.E.D.

In Lemma 4.1, we  resort to the results on the almost periodic
functions presented in Ref.[15]. Temporal behavior of the almost
periodic functions is rather subtle [15,31,32] and, in general,
requires separate careful analysis. Here we
 have in mind the cases essentially described in Ref.[15] that do not apply to the few-particle systems
while $t_{\circ}$ appears to be of the order of "decoherence
time".

Lemma 4.1(i) implies $\rho^A_{\alpha} = \sum_m p_{\alpha m} \vert
m\rangle_A\langle m\vert, \sum_{m} p_{\alpha m} =
1,\forall{\alpha}$, and therefore the first term in eq.(23) in the
formal limit $t_{\circ} \to \infty$:

\begin{equation}
 \sum_{\alpha,m} \vert
b_\alpha\vert^2 p_{\alpha m} \vert \alpha\rangle_O\langle \alpha
\vert \otimes \vert m\rangle_A\langle m\vert,
\end{equation}

\noindent which is the so-called "classical-classical" state with
zero two-way discord [33], $D^{\leftrightarrow}(O\vert A) = 0$,
i.e. without quantum correlations. Notice that the states $\vert
\alpha\rangle_O$ diagonalize $H_{int}$, while, in general, this is
not the case with the  $\vert m\rangle_A$ states in eq.(28).

With the use of notation of Lemma 4.1, eq.(23) gives rise to:

\begin{equation}
 \lim_{t_{\circ} \to\infty} \rho^O = \lim_{t_{\circ}
\to\infty} tr_A \sigma = \sum_{\alpha} \vert b_\alpha \vert^2
\vert \alpha \rangle_O
 \langle \alpha \vert,
\rho^A = tr_O \sigma = \sum_{\alpha} \vert b_\alpha \vert^2
\rho^A_{\alpha},
\end{equation}

\noindent which are the states at the observer's disposal; only
for the few-particle systems, the observer may have access to the
total system's state eq.(23).

\bigskip

{\bf 4.2 Unique pointer basis}

\bigskip

Orthogonality of the $\rho^A_{\alpha}$s, Lemma 4.1(i), implies
that they have orthogonal support. Then from eq.(29), the mutual
information $I(O:A)$ [25] can  easily be calculated [in the formal
limit $t_{\circ} \to \infty$]:

\begin{equation}
I(O:A) = S(\rho^A) - \sum_{\alpha} \vert b_{\alpha}\vert^2 S(
\rho^A_{\alpha}) = H(O),
\end{equation}

\noindent where $S(\rho) = - tr \rho \ln \rho$ is the von Neumann
entropy and $H(O) = -\sum_{\alpha} b_{\alpha} \ln b_{\alpha}$ is
the Shannon entropy of the $O$'s state; the last equality in
eq.(30) is a direct consequence of Theorem 11.10 in Ref. [25]. In
the context of decoherence theory, eq.(30) exhibits that the
environment carries  classically distinguishable records about the
open system's states $\vert \alpha\rangle_O$. In this sense, we
can see that the quantum environment is performing a measurement
on the open system [34].

For the pure state case, e.g. eq.(8), of the total system arises a
problem as the Schmidt form of the state need not be unique. This
happens only if $\rho_O$, eq.(29), has degenerate spectrum, $\vert
b_{\alpha}\vert^2$, as a consequence of the choice of the initial
state of the $O$ system. Then eq.(30) might simultaneously apply
to mutually non-commuting observables--this is known as the
"preferred-basis problem" [13]. In the remainder of this section
we show that this is not the case for the mixed state $\sigma$,
eq.(23). Thus we learn an important technical lesson: even a tiny
mixedness in a bipartite many-particle system can remove the
ambiguity known for the Schmidt form of pure states of the total
system.

From eq.(23) it follows for an alternative basis $\vert
\nu\rangle_O$: $\sigma = \sum_{\nu,\nu'} \vert \nu\rangle_O\langle
\nu'\vert \otimes R^A_{\nu\nu'}$; $R^A_{\nu\nu'} =
\sum_{\alpha,\alpha'} b_{\alpha} b_{\alpha'}^{\ast} c_{\alpha\nu}
c_{\alpha'\nu'}^{\ast} \rho^A_{\alpha\alpha'}$ and $c_{\alpha\nu}
= _O\langle \nu\vert\alpha\rangle_O$. Due to Lemma 4.1(i),
$\rho^A_{\alpha\alpha'} \rho^A_{\alpha''\alpha'''} =
\delta_{\alpha'\alpha''} \rho^A_{\alpha\alpha'}
\rho^A_{\alpha'\alpha'''}$ so it follows:

\begin{eqnarray}
 && tr R^A_{\nu\nu'} = \sum_{\alpha} \vert
b_{\alpha}\vert^2 c_{\alpha\nu} c_{\alpha\nu'}^{\ast}\quad =\quad
_O\langle \nu\vert \left(\sum_{\alpha} \vert b_{\alpha}\vert^2
\vert \alpha\rangle_O\langle \alpha\vert\right)\vert
\nu'\rangle_O,\\&& \nonumber (R^A_{\nu}R^A_{\nu'})_{\beta\beta'} =
\sum_{\alpha,\alpha'} b_{\alpha} b_{\alpha'}^{\ast} c_{\alpha\nu}
c_{\alpha'\nu'}^{\ast} \sum_{\alpha''} \vert b_{\alpha''}\vert^2
c_{\alpha''\nu}^{\ast} c_{\alpha''\nu'} (\rho^A_{\alpha\alpha''}
\rho^A_{\alpha''\alpha'})_{\beta\beta'} = \\&&
\sum_{\alpha,\alpha'} b_{\alpha} b_{\alpha'}^{\ast} c_{\alpha\nu}
c_{\alpha'\nu'}^{\ast} \quad _O\langle
\nu'\vert\left(\sum_{\alpha''} \vert b_{\alpha''}\vert^2
(\rho^A_{\alpha\alpha''}
\rho^A_{\alpha''\alpha'})_{\beta\beta'}\vert
\alpha''\rangle_O\langle \alpha''\vert\right)\vert \nu\rangle_O.
\end{eqnarray}

 For degenerate spectrum of $\rho_O$, eq.(29), that is for at
least two equal  $\vert b_{\alpha}\vert^2$s, we can choose an
alternative basis for which $tr R^A_{\nu\nu'} = 0,
\forall{\nu\neq\nu'}$, when the point (ii) of Lemma 4.1 applies to
the new basis $\vert \nu\rangle_O$. E.g. for $\vert b_1\vert^2 =
\vert b_2 \vert^2$, eq.(31) is fulfilled for the orthonormalized
basis $\{\vert \nu_1\rangle_O,\vert \nu_2\rangle_O, \vert
\alpha\rangle_O, \alpha = 3,4,5,\dots\}$ for $\vert \nu_i
\rangle_O = \sum_{\alpha=1}^2 c_{i\alpha} \vert \alpha\rangle_O,
i=1,2$. However, as long as  $\langle \nu\vert\ast\vert\nu'\rangle
\approx 0$ in eq.(31), the matrix element in eq.(32) $\langle
\nu'\vert\ast\vert\nu\rangle \not\approx 0$, and {\it vice versa}.

On the other hand, Lemma 4.1(i) requires that the condition
$(R^A_{\nu}R^A_{\nu'})_{\beta\beta'} \approx 0$ is satisfied for
all combinations of the indices $\nu\neq\nu', \beta, \beta'$ as
well as for most of the large values of $t_{\circ}$. As Lemma 4.1
is concerned with the many-particle systems, there are thus a huge
number of equations that should be satisfied simultaneously. E.g.
for fixed $\nu,\nu'$, and for $n$ qubits in the $A$ system. there
are $2^{n-1}(2^n+1)$ equations that should be simultaneously
satisfied for most of the large values of $t_0$.  These arguments
yield: there is not any alternative basis $|\nu\rangle_{O}$ with
$c_{\alpha\nu} \neq \delta_{\alpha\nu}$ for which both points, (i)
and (ii) of Lemma 4.1, could be valid for every combination of the
indices $\nu\neq\nu', \beta$ and $\beta'$ as well as for most of
the large values of $t_{\circ}$ of local time of a many-particle
$O+A$ system.

 Now Lemma 4.1 and eqs.(23)-(32) {\it uniquely} determine the
 pointer basis $\{\vert \alpha\rangle_O\}$ as well as the
"superselection sectors" $P^O_{\alpha}$ and the "pointer
observable" $A^O = \sum_{\alpha} a_{\alpha} P^O_{\alpha}$.
 In
contrast to the case of the pure state of the total system,
degeneracy in the $O$ system's state eq.(29) does {\it not} give
rise to ambiguity in regard of what is "measured" in the
local-time scheme. This conclusion holds independently of the
initial state of the $O$ system as well as of the
interaction-energy spectrum and the number of particles in the $A$
system, $N$ [except that $N \gg 1$]. This brings the {\it main
result} of this paper:

\smallskip

\noindent A bipartition of a {\it closed}, finite-dimensional or
infinite-dimensional, {\it many-particle} system can exhibit
behavior that is characteristic for  {\it open} systems. Neither
"objective wave packet collapse" nor the environmental influence
is required.

\bigskip

{\bf 4.3 Analysis of quantum measurement}

\bigskip

Many-particle (macroscopic) systems are always in inevitable
interaction with their environments. In this section we extend the
analysis of the previous sections to the case of measurement, when
the $A$ system is the "apparatus" and we introduce the apparatus'
environment ($E$), which does not interact with the $O$ system.
This scenario clearly distinguishes the two standard "phases" in
quantum measurement: In the standard measurement theory,
interaction of the $O$ and $A$ systems gives rise to
"premeasurement" [13-15], i.e. to formation of quantum
entanglement in the $O+A$ system. The second phase of measurement
is decoherence of the apparatus that is induced by the environment
[13,14].

In the local-time scheme, the two phases of measurement are
clearly distinguished. According to the point (a) of Section 2.2,
they refer to different many-particle systems, $O+A$ and $O+A+E$
and their related local times, respectively. Regarding
premeasurement, in an instant $t_{\circ}$ of local time for the
$O+A$ system, the total system's state is (approximately) given in
a Schmidt canonical form, $\sum_i b_{\alpha} \vert
\alpha\rangle_O\vert \alpha \rangle_A$. Now, in the local-time
scheme, the second phase  considers another, newly formed
many-particle system, $O+A+E$, which dynamically evolves in
accordance with {\it its own} local time and re-sets its own time
to the instant $t=0$.

The correlation between the $O$ and $A$ systems is preserved by
the environment [13-17] and carries the information that
constitutes the measurement performed on $O$ by the $A$ system. Of
course, this requires robustness of the apparatus' states $\vert
\alpha\rangle_A$ as the very basic requirement of successful
measurement [13-17]. For the tripartite system $O + A + E$, the
Schr\" odinger dynamics gives:

\begin{equation}
U(t) \sum_{\alpha,j} b_{\alpha} d_j \vert \alpha\rangle_O \vert
\alpha \rangle_A \vert j \rangle_E = \sum_{\alpha,j} b_{\alpha}
d_j \exp(-\imath t h_{\alpha j}/\hbar) \vert \alpha\rangle_O \vert
\alpha \rangle_A \vert j \rangle_E,
\end{equation}

\noindent while assuming strong interaction between the $A$ and
$E$ systems; without loss of generality, we can ignore degeneracy
in the interaction between $A$ and $E$. Then the state eq.(23)
takes the following form:

\begin{equation}
\sigma = \sum_{\alpha} \vert b_{\alpha} \vert^2 \vert
\alpha\rangle_O\langle \alpha \vert \otimes \vert
\alpha\rangle_A\langle \alpha \vert \otimes \rho^E_{\alpha} +
\sum_{\alpha\neq \alpha'} b_\alpha b_{\alpha'}^{\ast} \vert
\alpha\rangle_O\langle \alpha' \vert \otimes \vert
\alpha\rangle_A\langle \alpha' \vert \otimes
\rho^E_{\alpha\alpha'}.
\end{equation}

It can be easily shown that the operators $\rho^E_{\alpha}$ and
$\rho^E_{\alpha\alpha'}$ are exactly of the form of eqs.(24) and
(25), respectively, and Lemma 4.1 applies. Consequently, the
conclusions are analogous: the environment monitors the composite
system $O+A$, while not affecting the correlations formed in
premeasurement in the $O+A$ system.

Bearing Lemma 4.1 in mind, now eq.(34) implies:

\begin{equation}
\rho^{O+A} = tr_E \sigma = \sum_{\alpha} \vert b_{\alpha} \vert^2
\vert \alpha\rangle_O\langle \alpha \vert \otimes \vert
\alpha\rangle_A\langle \alpha \vert, \rho^O = tr_A \rho^{O+A} =
\sum_{\alpha} \vert b_{\alpha}\vert^2 \vert \alpha\rangle_O\langle
\alpha \vert.
\end{equation}

 So the local-time scheme
straightforwardly reproduces the basic result eq.(35) of the
environmental influence on the "object+apparatus" system
[13-15,17].

\bigskip

{\bf 5. Some models of quantum decoherence and measurement}

\bigskip

In this section we analyze some relevant analytically-solvable
models described by  "pure decoherence" [18,34,35] Hamiltonian
eq.(20). We observe technical simplicity and clarity of the
local-time scheme. In accordance with Section 3, we choose the
highest possible value for $\Delta t$ and the smallest possible
value for $\lambda$. Coarse graining of the pointer observable
($A^O$) values is considered without a change of values of the
$\Delta t$ and $\lambda$ parameters; for motivation see Section 6.

\bigskip

{\bf 5.1 A pair of spin-1/2 particles}

\bigskip

Consider a pair of spin-1/2 particles [qubits] and interaction
$H_{int} = C S_{1z} S_{2z}$. This is a separable interaction [30],
cf. eq.(20), with the eigenstates $\vert ++\rangle, \vert
+-\rangle, \vert -+\rangle, \vert --\rangle$ and eigenvalues (in
the units of the Planck constant, $\hbar = 1$) $h_{++} = C/4 =
h_{--}, h_{+-} = -C/4 = h_{-+}$, while the ground energy $E_g =
-C/4$.

Let us consider the mixed state eq.(23) for this case:

\( \rho_{2+} =
\begin{pmatrix}
   \vert d_+\vert^2  & d_+ d_{-}^{\ast} e^{{-\imath t_{\circ}C\over 2} - {C^2\over 16\lambda}}  \\
   d_+^{\ast} d_- e^{{\imath t_{\circ}C\over 2} - {C^2\over 16\lambda}} & \vert d_-\vert^2
  \end{pmatrix}
\)

\noindent and

\( \rho_{2-} =
\begin{pmatrix}
   \vert d_+\vert^2  & d_+ d_{-}^{\ast} e^{{\imath t_{\circ}C\over 2} - {C^2\over 16\lambda}}  \\
   d_+^{\ast} d_- e^{{-\imath t_{\circ}C\over 2} - {C^2\over 16\lambda}} & \vert d_-\vert^2
  \end{pmatrix}
\)

 \noindent while

\( \rho_{2+-} =
\begin{pmatrix}
   \vert d_+\vert^2 e^{-\imath t_{\circ}C/2}e^{-C^2/16\lambda}  & d_+ d_{-}^{\ast}   \\
   d_+^{\ast} d_-  & \vert d_-\vert^2 e^{\imath t_{\circ}C/2}e^{-C^2/16\lambda}
  \end{pmatrix}.
\)

Now it easily follows: 
\begin{eqnarray}
&\nonumber& (\rho^{(2)}_{+} \rho^{(2)}_{-})_{++} = \vert
d_+\vert^2 e^{-\imath t_{\circ}C/2} [\vert d_+\vert^2 e^{\imath
t_{\circ}C/2} + \vert d_-\vert^2 e^{-\imath t_{\circ}C/2}
e^{-C^2/8\lambda}]
\\&&
tr_2 \rho^{(2)}_{+-} = e^{-C^2/16\lambda} [\vert d_+\vert^2
e^{-\imath t_{\circ}C/2} + \vert d_-\vert^2 e^{\imath
t_{\circ}C/2}] = e^{-C^2/16\lambda} \cos{C t_{\circ} \over 2}
\end{eqnarray}

\noindent i.e. Lemma 4.1 is not fulfilled for this case.

 The choice $d_{\pm} = 2^{-1/2}$ satisfies the condition
 $\langle H_{int} \rangle_{t=0} = 0$ and $\Delta H_{int} = C/4 = - E_g$.
 Then there is the unique time bound, $\tau_{min}/2 = \pi/C$, and for $C=1$ we
can choose $\Delta t = 3$ and $\lambda = 1$, with the very well
satisfied equality eq.(14). Thus $\exp[-C^2/16\lambda] =
\exp(-1/16) \approx 0.939$ and the small off-diagonal term,
$\exp[-(h_{++}-h_{--})^2/4\lambda] = \exp(-1/16) \approx 0.939$.
Therefore

\begin{equation}
\sigma \approx \vert \Psi\rangle\langle \Psi\vert,
\end{equation}

\noindent with the error less than $0.062$, [the error decreases
with the increase of $\lambda$], and with  $\vert \Psi\rangle =
[c_+\exp(-\imath t_{\circ} /4) \vert ++\rangle + c_-\exp(-\imath
t_{\circ} /4) \vert --\rangle + c_+\exp(\imath t_{\circ} /4) \vert
+-\rangle + c_-\exp(\imath t_{\circ} /4) \vert
-+\rangle]/\sqrt{2}$--that is eq.(8) for this case; $\vert
c_+\vert^2 + \vert c_-\vert^2 = 1$. So quantum coherence is very
high in the local-time scheme for   "microscopic" systems--and is
analogous to the approximate isolation (and coherence) of the
microscopic systems in the context of open quantum systems  as
anticipated in Section 3.2.

\bigskip

{\bf 5.2 Four spin-1/2 particles: a case study}

\bigskip

A spin-1/2 system is in interaction with mutually noninteracting
spin-1/2 systems [qubits]: $H_{int} = S_{1z} (S_{2z} + S_{3z} +
S_{4z})$. The system $2+3+4$ is the $1$ system's environment. The
interaction is separable [30], eq.(20), and the eigenstates and
eigenvalues [in the units $\hbar = 1$] can be denoted $\vert \pm i
\rangle$ and $h_{\pm i}$, respectively. The indices $\pm$ refer to
the $1$ system, while the index $i=1,2,...,8$ denotes the set of
mutually orthogonal tensor-product states, $\vert m_2 n_3
p_4\rangle, m,n,p = \pm$, which constitute an orthonormalized
basis for the $2+3+4$ system. The eigenvalues $h_{\pm\beta}$ and
degeneracies $g_{\beta}$ are as follows: $h_{\pm1} = \pm 3/4 =
h_{\mp 4}$, $h_{\pm 2} = \pm 1/4 = h_{\mp 3}$ and $g_1 = 1 = g_4$,
while $g_2 = 3 = g_3$.

Again we choose  $\langle H_{int} \rangle = 0$ that is satisfied
e.g. with equal distribution of eigenstates  $\vert m_2 n_3
p_4\rangle$ for the initial environment's state, $ 8^{-1/2},
\forall{i=1-8}$. Then, bearing in mind the degeneracies, with the
use of notation of eqs.(21), (22), $\vert d_1\vert^2 = 1/8 = \vert
d_4\vert^2$ and $\vert d_2\vert^2 = 3/8 = \vert d_3\vert^2$. For
arbitrary initial state of the $1$ system, $\tau_{min}/2 =
\pi/4\Delta H_{int} = \pi/4(\langle H_{int}\rangle - E_g) =
\pi/3$. Hence we can choose $\Delta t = 1$ and $\lambda = 2$ in
order to provide a very good approximation for eq.(14). This gives
e.g.: 
\begin{eqnarray}
&\nonumber&  (\rho^{(234)}_{ +} \rho^{(234)}_{ -})_{11} = \vert
d_1\vert^2 \exp(-3\imath t_{\circ}/2) \quad [\vert d_1\vert^2
\exp(3\imath t_{\circ}/2) +\nonumber\\&& \vert d_2\vert^2
\exp(\imath t_{\circ}/2-1/16) + \vert d_3\vert^2  \exp(-\imath
t_{\circ}/2-1/4) + \vert d_4\vert^2 \exp(-3\imath
t_{\circ}/2-9/16)];
 \nonumber\\&&
tr_{234} \rho^{(234)}_{+-} = \vert d_1\vert^2  \exp(-3\imath
t_{\circ}/2-9/32) + \vert d_2\vert^2  \exp(-\imath
t_{\circ}/2-1/32) +\nonumber\\&& \vert d_3\vert^2 \exp(\imath
t_{\circ}/2-1/32) + \vert d_4\vert^2  \exp(3\imath
t_{\circ}/2-9/32) =\nonumber
\\&& {1 \over 4} \cos ({3t_{\circ} \over 2}) \exp(-9/32) + {3\over
4} \cos({t_{\circ}\over 2}) \exp(-1/32).
\end{eqnarray}

Needless to say, due to  small number of terms in the sums in
eq.(38), Lemma 4.1 is not satisfied. Nevertheless, comparison of
eqs.(36) and (38) clearly exhibits that increase in the size of
the environment gives better satisfied Lemma 4.1. Both traces in
eqs.(36) and (38) are periodic functions [with the periods
approximately $2\pi$ and $4\pi$] and the increase in the number of
terms in the sum leads to the almost periodic functions, Lemma
4.1. Physically, eq.(38) reveals environment's periodic memory,
with small period, about the object's state--that is not a good
measurement or decoherence of the $1$ system.

The real exponential terms $\exp[-(h_{\alpha\beta} -
h_{\alpha'\beta'})^2/4\lambda]$ for the above set of energy
eigenvalues and for small $\lambda = 2$ have the smallest value
$\exp(-9/32) = 0.755$ and the largest value $\exp(-1/32) = 0.969$.
Compared to the previous model, there is less quantum coherence in
the total system.  So

\begin{equation}
\sigma \not\approx \vert \Psi'\rangle\langle \Psi' \vert
\end{equation}

\noindent where $\vert \Psi'\rangle =
\sum_{\alpha=\pm}\sum_{\beta=1}^4 b_{\alpha} d_{\beta}
\exp(-\imath t_{\circ} h_{\alpha\beta}) \vert \alpha\beta \rangle$
is the standard state eq.(8). The fidelity satisfies $0.869 =
\sqrt{0.755} < \mathcal{F} = \sqrt{\langle \Psi' \vert \sigma
\vert \Psi'\rangle} < \sqrt{0.969} = 0.984$. After a
straightforward but lengthy computation, the exact fidelity
amounts to $0.894$. Decrease of quantum coherence relative to the
model of Section 5.1 supports and illustrates the general notions
provided in Section 3.2: the larger the environment the less
quantum coherence  in the total system.

\bigskip

{\bf 5.3 Decoherence and measurement of a single qubit: the qubit
environment}

\bigskip

We consider the well-studied, analytically solvable model of
"decoherence of a single qubit" [15]. This also models the
Stern-Gerlach experiment, if the environment is modelled as the
set of molecules in the plate that can be either decayed or
non-decayed by the atoms caught by the plate.

The interaction Hamiltonian for the pair $O+A$, where the $O$
system is the single qubit is separable:

\begin{equation}
H_{int} = (a_+\vert +\rangle_O\langle +\vert + a_- \vert
-\rangle_O\langle -\vert) \otimes \sum_{k=1}^N g_k (\vert +
\rangle_{Ak}\langle +\vert - \vert -\rangle_{Ak}\langle -\vert)
\Pi_{k\neq k'}I_{k'},
\end{equation}

\noindent with $a_+ = - a_- = 1$ and with $N \gg 1$.

Initial state $\vert \Psi\rangle = (a\vert +\rangle_O + b\vert
-\rangle_O) \Pi_{k=1}^N (a_k \vert +\rangle_{Ak} + b_k \vert
-\rangle_{Ak})$ gives for an instant of time [15]:

\begin{equation}
\vert \Psi (t)\rangle = a \vert +\rangle_O \vert
\chi_+(t)\rangle_A + b \vert -\rangle_O \vert
\chi_-(t)\rangle_{A},
\end{equation}

\noindent where [for $\hbar=1$]

\begin{equation}
\vert \chi_{\pm}(t)\rangle_A = \Pi_{k=1}^N (a_k e^{-\imath a_{\pm}
g_kt}\vert +\rangle_{Ak} + b_k e^{\imath a_{\pm}g_kt}\vert
-\rangle_{Ak}).
\end{equation}

Eq.(42) can be written as:

\begin{equation}
\vert \chi_{\pm}(t)\rangle_A = \sum_{j_1 \dots j_N = \pm} c_{j_1}
\dots c_{j_N}  \Pi_{k=1}^N e^{-\imath ta_{\pm}  g_k\alpha_k} \vert
j_k\rangle_A
\end{equation}

\noindent where $\alpha_k = \nu_k-\mu_k$ and $\nu, \mu = 0,1$ with
the following rule: if $j_k = +$, then $\nu_k = 1$ and $\mu_k =
0$, while for $j_k = -$,  $\nu_k = 0$ and $\mu_k = 1$, with
independent constants for different indices $k$.

Setting $ \Pi_{k=1}^N e^{-\imath ta_{\pm} g_k\alpha_k} \vert
j_k\rangle_A =  e^{-\imath ta_{\pm} \sum_k g_k\alpha_k}
\Pi_{k=1}^N  \vert j_k\rangle_A$, the  mixed state eq.(9):

\begin{equation}
\sigma = \vert a\vert^2 \vert +\rangle_O\langle +\vert \otimes
\rho^A_{+} + \vert b\vert^2 \vert -\rangle_O\langle -\vert \otimes
\rho^A_{-} +  ab^{\ast} \vert +\rangle_O\langle -\vert \otimes
\rho^A_{+-} + a^{\ast} b \vert -\rangle_O\langle +\vert \otimes
\rho^A_{-+}.
\end{equation}

In eq.(44):

\begin{eqnarray}
&\nonumber& \rho^A_{\pm}(t_{\circ}) = \sum_{j_1\dots j'_N}
c_{j_1}\dots c^{\ast}_{j'_N} e^{-\imath t_{\circ} a_{\pm} \sum_k
g_k(\alpha_k - \alpha'_k)} e^{-(a_{\pm} \sum_k g_k(\alpha_k -
\alpha'_k))^2/4\lambda} \Pi_k \vert j_k\rangle_A\langle j'_k\vert
\nonumber
\\&&
\rho^A_{+-}(t_{\circ}) = \sum_{j_1\dots j'_N} c_{j_1}\dots
c^{\ast}_{j'_N} e^{-\imath t_{\circ} \sum_k g_k (a_+\alpha_k -
a_-\alpha'_k)} \times \nonumber\\&& e^{-(\sum_k g_k (a_+\alpha_k -
a_-\alpha'_k))^2/4\lambda} \Pi_k \vert j_k\rangle_A\langle
j'_k\vert
\end{eqnarray}

\noindent where $\alpha_k$ refers to $j_k$ and $\alpha'_{k}$ to
$j'_k$.

From eq.(45): 
\begin{eqnarray}
&\nonumber& \rho^A_{+}(t_{\circ})\rho^A_{-}(t_{\circ}) =
\sum_{j_1\dots j_N, j'_1\dots j'_N} c_{j_1}\dots c_{j_N}
c^{\ast}_{j'_1}\dots c^{\ast}_{j'_N} e^{-\imath t_{\circ} \sum_k
g_k(a_+\alpha_k - a_-\alpha'_k)} \Pi_k\vert j_k\rangle_A\langle
j'_k\vert  \times \nonumber\\&& \left(\sum_{j''_1\dots j''_N}
\vert c_{j''_1} \vert^2 \dots \vert c_{j''_N} \vert^2 e^{-\imath
t_{\circ} (a_- - a_+)\sum_k g_k\alpha''_k}  e^{-
{(a_+\sum_kg_k(\alpha_k - \alpha''_k))^2 + (a_-\sum g_k(\alpha''_k
- \alpha'_k))^2 \over4\lambda}}\right) \nonumber\\&& tr_A
\rho^A_{+-} = \sum_{j_1 \dots j_N} \vert c_{j_1}\vert^2 \dots
\vert c_{j_N}\vert^2 e^{-\imath t_{\circ}(a_+ - a_-)\sum_k
g_k\alpha_k} e^{-((a_+ - a_-)\sum_k g_k\alpha_k)^2/4\lambda}.
\end{eqnarray}

The term in the parenthesis and the trace $tr_A\rho^A_{+-}$ are of
the form of the $\chi$-function defined in the proof of Lemma
4.1(i)--see below eq.(26). Therefore, Lemma 4.1 applies for the
case studied: symbolically, $\lim_{t_{\circ} \to \infty}
\rho^A_{+}\rho^A_{-} = 0$ and
$\lim_{t_{\circ}\to\infty}tr_A\rho^A_{+-} = 0$ for $N \gg 1$.

In order to compare with Zurek's [15], we deal with the random
values for $g_k \in (0,1)$ and $\vert a_k\vert \approx \vert
b_k\vert, \forall{k}$; the latter gives rise to $\langle
H_{int}\rangle \approx 0$. It easily follows $\Delta H_{int} =
\sqrt{\sum_k g_k^2}$. For randomly chosen $g_k = k/N$ and with
equal probability $1/N$ for every $g_k$, $\Delta H_{int} =
\sqrt{N^{-3}\sum_{k=1}^N k^2} = 3^{-1/2} > - E_g =
N^{-2}\sum_{k=1}^{N} k = 1/2$, for $N \gg 1$. Therefore
$\tau_{min}/2 = \pi/2 > 1.57$. So we choose $\Delta t = 1.56$ and
the smallest value $\lambda = 1$ that provide very good
approximation for eq.(14).

Exponential factors appearing in the $\rho^A_{+-}$ in eq.(45):

\begin{equation}
e^{-(\sum_k g_k (a_+\alpha_k \pm a_-\alpha'_k))^2/4\lambda} = e^{-
(\sum_k g_k (\alpha_k \pm \alpha'_k))^2/ 4} .
\end{equation}

Since $\max\{\alpha_k \pm \alpha'_k\} = 2$, the smallest
exponential factor is $\exp(-1/4) = 0.779$. All  other terms are
with the nominator in the exponent of the form $(\pm\sum_{k=1}^M
g_k \mp \sum_{k=M+1}^N g_k)^2 = (N^{-2}[\pm\sum_{k=1}^M k \mp
\sum_{k=M+1}^N k])^2$. Numerical estimates reveal that such terms
are not less than $0.94$. In order to compare with the model of
Section 5.2, we set $N=3$ [and placing $\lambda = 2$ and the
eigenvalues $\pm 1$ instead of $\pm 1/2$] and obtain similar
results. So we find that there is high quantum coherence for both
models of Section 5.2 and of this section.

Without further ado, let us consider the object's spectrum $a_i
\in \{-2,-1,1,2\}$--which can describe the four spin-1/2 particles
total-spin values; the $a_i$ values substitute  the above
$a_{\pm}$ values. For a pair of  values, e.g. $2$ and $-1$, and
for random $g_k$s (see above) while $N\gg 1$, the smallest
Gaussian factor $\exp[-(\sum_k g_k(2\alpha_k - \alpha'_k))^2/4] =
\exp(-9/16) \approx 0.57$, while the largest one amounts to
$\exp(-1/16) = 0.939$. For the pair $2, -2$, the smallest term
$\exp(-1) = 0.368$ while the largest amounts to $1$.
 Now consider the coarse graining of this spectrum by introducing
the new set of values, $a'_j \in \{-2,0,2\}$. For the values  $2$
and $0$, [with $\lambda = 1$], there is the unique value of
$\exp(-1/4) = 0.778$, while the  terms pertaining to the pair $2,
-2$,  remain intact.

So we obtain a rough idea about decrease of coherence due to the
coarse graining of spectrum of the pointer observable, and
consequently of energy in the composite system: the number of
large Gaussian terms decreases. Needless to say, due to the poor
spectrums, this is not possible for the microscopic objects of
Sections 5.1, 5.2 and eq.(41). In turn, we also realize: finer
measurements--e.g. of the spectrum $a_i$ instead of the
coarse-grained values $a'_j$--can in principle provide observation
of coherence, i.e. of quantum correlations in the total system.

\bigskip

{\bf 5.4 Position measurement}

\bigskip

The classic von Neumann's model [36] that implements the
Heisenberg's idea of  position measurement is described by strong
interaction $H_{int} = C x_O \otimes P_A$ between the
one-dimensional object $O$ and the apparatus $A$; the conjugate
momentum/position observables $p_O$ and $X_A$, respectively. The
model is readily generalized for measurement of any continuous
observable $Q_O$ as well as to the three dimensional models [37].
Similar results are obtained for the interaction $H_{int} = x_O
\otimes X_A$. For the collective position observable $X_A = \sum_j
\kappa_j x_{Aj}$, the object $O$ undergoes  quantum Brownian
motion [19] that does not depend on the strength of interaction.

Let us consider the composite system initially spatially contained
in the linear dimensions $[-L,L]$ and the initial state $\vert
\phi\rangle_O\vert\chi\rangle_A$ as a tensor product of two
wavepackets with the position and momentum spreads
$\sigma_{x_O}\equiv \sigma_1$ and $\sigma_{P_A} \equiv \sigma_2$,
while for convenience $\langle H_{int}\rangle = 0$. For the
analogous interval for the apparatus momentum $[-P,P]$ the ground
energy $E_g = - LP \ll 1$. If the spreads $\sigma_{1} \sim 1$ and
$\sigma_2 \sim 1$, then [in the units $\hbar=1$ and for $C=1$]
$\tau_{min}/2 = \max\{\pi/4\sigma_1\sigma_2, \pi/4LP\} = \pi/4$,
while $\Delta t = 0.78$ and $\lambda = 3$  well satisfy eq.(14).

Then the  state eq.(23) reads:

\begin{equation}
\sigma = \int dx dx' \vert x\rangle_O\langle x'\vert \otimes
\rho_A(x,x'),
\end{equation}

\noindent with 
\begin{equation}
\rho_A(x,x') = \int dP dP' \phi(x)\phi^{\ast}(x')
\chi(P)\chi^{\ast}(P') \exp(-\imath t_{\circ} (xP - x'P'))
\exp(-{(xP-x'P')^2\over 12}) \vert P\rangle_A\langle P'\vert.
\end{equation}

From eq.(49) one easily obtains validity of Lemma 4.1,  due to
direct applicability of the Riemann-Lebesgue lemma, cf. e.g.
Proposition 5.2.1 in Ref.[15], in our case:
$\lim_{t_{\circ}\to\infty}\int dP \vert \chi(P)\vert^2
\exp(-\imath t_{\circ} (x'-x)P)
\exp\{-[(xP'-xP)^2+(x'P-x'P'')^2]/12\} = 0$ for $x\neq x'$.

The fidelity $\sqrt{\int dx dx' dP dP' \vert \phi(x)\vert^2 \vert
\phi(x') \vert^2 \vert\chi(P) \vert^2 \vert\chi(P') \vert^2
\exp(-(xP-x'P')^2/12)}$ reveals very high coherence for the
object's state--there are plenty of close $x$ and $x'$.
Nevertheless, there are still very small values for the Gaussian
factors for which $\vert x - x'\vert \gg 12$--that is well within
the chosen domain of $L \gg 1$.

Coarse graining of the pointer-observable $x_O$ continuous
spectrum (while keeping the parameter $\lambda$ fixed) reduces the
number of the Gaussian terms, which almost equal 1. If the width
of the spatial interval is $\Delta x$, then one can choose the
wavepackets with the spread $\Delta x$ as the approximate
(non-orthogonal) normalizable  "pointer basis" states. Formally,
for a set of approximately orthogonal minimum-uncertainty (the
"coherent") states $\vert \psi_{ij}\rangle_O$, such that
$_O\langle\psi_{ij}\vert \psi_{i'j'}\rangle_O \approx \delta_{ii'}
\delta_{jj'}$, one obtains $_O\langle\psi_{ij}\vert x_O\vert
\psi_{i'j'}\rangle_O \approx x_i \delta_{ii'} \delta_{jj'}$. Then
the exact interaction is almost diagonal for the $\vert
\psi_{ij}\rangle_O$ states: $_O\langle \psi_{ij}\vert H_{int}\vert
\psi_{i'j'}\rangle_O \approx 0, \forall{i\neq i', j,j'}$.
Furthermore, the unitary operator generated by the interaction is
also almost diagonalizable for these states. The proof reduces to
computing the $_O\langle \psi_{ij}\vert
x^n_O\vert\psi_{i'j'}\rangle_O$ terms. For $\psi_{ij}(x) =
(2\pi\sigma_i)^{-1/2} \exp(-(x-x_i)^2/2\sigma_i^2 + \imath xp_j)$:

\begin{equation}
_O\langle \psi_{ij}\vert x_O^n\vert \psi_{i'j'}\rangle_O =
(2\pi\sigma_i\sigma_{i'})^{-1} \exp(-(x_i -
x_{i'})^2/2(\sigma_i^2+\sigma_{i'}^2)) \quad \mathcal{I}_n
\end{equation}

\noindent where $\mathcal{I}_n = \int_{-\infty}^{\infty} dx x^n
\exp(-(x-x_{\circ})^2/2\sigma^2 - \imath x(p_j-p_{j'}))$;
$\sigma^2 = \sigma_i^2\sigma_{i'}^2/(\sigma_i^2+\sigma_{i'}^2)$
and $x_{\circ} =
(x_{i'}\sigma_i^2+x_{i}\sigma_{i'}^2)/(\sigma_i^2+\sigma_{i'}^2$).
The Gaussian term in eq.(50) proves the claim: $\Vert _O\langle
\psi_{ij}\vert U(t_{\circ})\vert \psi_{i'j'}\rangle_O\Vert \propto
\exp(-(x_i - x_{i'})^2/4) \ll 1$.

Since $\sum_{i,j} \vert \psi_{ij}\rangle_O\langle \psi_{ij}\vert <
I$, there are plenty of "coherent states" in the vicinity of every
$\vert \psi_{ij}\rangle_O$ that contribute to degeneracy of the
interaction. Hence for the set of the values $x_i$ (out of the
continuous set of the position values $x \in (-\infty,\infty)$)
one obtains substantial decrease of the Gaussian factors, while
the coherent states $\vert \psi_{ij}\rangle_O$ constitute a set of
approximate pointer basis states for the exact continuous pointer
observable $x_O$. The more rigorous methods [36,38]  give rise to
redefinition of the exact pointer observable and interaction and
hence of the $\Delta t$ and $\lambda$ parameters that we are not
interested in--see Section 6.

\bigskip

{\bf 5.5 Walls-Collet-Milburn measurement model}

\bigskip

The open system $O$ and the apparatus $A$ are taken to be harmonic
oscillators defined by the respective annihilation operators, $a$
and $b$ (the modes) and with the separable interaction [39]:

\begin{equation}
H_{OA} = {\hbar \over 2} a^{\dag}a (\epsilon^{\ast} b + \epsilon
b^{\dag}).
\end{equation}

There is also the apparatus environment $E$, which is a thermal
bath of harmonic oscillators with the interaction

\begin{equation}
H_{AE} = b \sum_j \kappa_j^{\ast} c_j^{\dag} + b^{\dag} \sum_j
\kappa_j c_j
\end{equation}

\noindent with the environmental annihilation operators (the
modes) $c_j$. The thermal bath can be "purified" and appears as a
subsystem of a larger system, which is initially in a pure state
that we are concerned with, cf. eq.(8), and will continue to be
denoted by $E$.

According to Section 4.3, eq.(51) describes pre-measurement,
Section 4.3, that gives rise to the final state of the $O+A$
system [19]:

\begin{equation}
\vert \Psi(t)\rangle_{OA} = \sum_n c_n \vert n\rangle_O\vert
n\epsilon t/2\rangle_A,
\end{equation}

\noindent where $a^{\dag}a\vert n\rangle_O = n \vert n\rangle_O$
and the apparatus states are "coherent states" [the minimum
uncertainty Gaussian states]. Setting $t = t_{\circ} \to \infty$,
the apparatus states are approximately orthogonal [14] and in the
instant of time $t_{\circ}$,  pre-measurement is complete.

The second phase of the measurement, cf. Section 4.3, is described
by the interaction eq.(52). By following Ref.[40], the interaction
eq.(52)
 is obtained via the so-called rotating-wave approximation
[19,20] that reveals the Schr\" odinger-picture, original
interaction to be of the separable form [39,40]:

\begin{equation}
H_{AE} = X_A [\sum_j \kappa_j^{\ast} c_j + \sum_j \kappa_j
c_j^{\dag}],
\end{equation}

\noindent where $X_A$ is the apparatus position observable.
Eq.(54) is of interest within the local-time scheme.

Eq.(54) is actually the model considered in Section 5.4: The
environment $E$ measures the apparatus' position $X_A$. So we
conclude that the second phase of the measurement--according to
Section 4.3--is an (almost ideal) "non-demolition"  measurement
[15,30] that distinguishes the $X_A$ observable as the pointer
observable with the approximate pointer basis $\vert n\epsilon
t/2\rangle_A$ for the apparatus. Needless to say, the object's
exact pointer observable is $a^{\dag}a$ and the exact pointer
basis states $\vert n\rangle_O$. As in eq.(35), the related
density matrices:

\begin{equation}
\rho_{O+A} = \sum_n \vert c_n\vert^2 \vert n\rangle_O\langle
n\vert\otimes\vert n\epsilon t/2\rangle_A\langle n\epsilon
t/2\vert, \rho_O = \sum_n \vert c_n\vert^2 \vert n\rangle_O\langle
n\vert.
\end{equation}

\bigskip

{\bf 6. Discussion}

\bigskip

In local-time scheme, the few-particle systems sustain high
quantum coherence. However, for bipartition of a many-particle
closed system we obtain effects that are characteristic for open
systems, without a need for the state collapse (reduction) or
environmental influence. Within the local-time scheme, "local
system" and "local operations" are defined via the set of local
time in a closed system. If certain pair interactions are of
similar strength, then the composite system can be subject to the
unique time, cf. eq.(5). The recipe for determining the local time
is conceptually rather simple as everything is written in the
total system's Hamiltonian: the degrees of freedom that are
relatively strongly coupled and (approximately) unitary evolve in
time constitute a subsystem, i.e. a "local system" that is defined
by its own local time that flows differently than for some other
local systems.  Those findings come from the macroscopic domain in
the context of the full quantum mechanical analysis. As distinct
from the Copehnagen interpretation, the local time scheme does not
assume or require "classical apparatus".

It is remarkable that the local-time scheme is technically simple.
It straightforwardly reproduces (Section 5) some basic results of
the standard decoherence and measurement theory. Amount of quantum
coherence in the total system depends on the system's state that
is reflected by the values of
 the  $\Delta t$ and $\lambda$ parameters. On the other hand,
coarse graining of the energy- and/or of the
pointer-observable-spectrum gives rise to a decrease of quantum
coherence as it is found in some other contexts [36,38,41-44].

In the context of our considerations, operational approach to
coarse graining [36,38,41-44] requires a change in the values of
the parameters $\Delta t$ and $\lambda$ and therefore in the time
bound $\tau_{min}$, Section 3.1. In the example of the microlocal
analysis [38], one introduces  quasi-projectors  and thus
redefines the position observable $x$ and consequently the
interaction considered in Section 5.4. The introduction of the new
sets of eigenvalues and (approximate) eigenspaces inevitably gives
rise to a change in the bound $\tau_{min}$--as it can be easily
shown. Not doing so, as we can see in Section 5, highlights the
observation of Section 3.2, that refining the measurement, i.e.
operational accessibility of the exact, "microscopic",
eigenvalues, can in principle give rise to observation of quantum
effects in the many-particle systems.

The local time scheme is easily adapted to reproduce the basic
assumptions of  diverse approaches to quantum foundations
involving emergent, relational, and fundamental time. First, the
scheme admits considerations (interpretation) that physical time
is {\it emergent}, i.e. not physically fundamental. To this end,
Time is a construct from the fundamental quantum dynamics, e.g.
presented by eq.(5). In this scheme  spacetime quantization may be
undefinable. Second, the Local Time Scheme provides  Relational
Character [45] of Common Local Time for interacting particles
(subsystems), cf. the point (b) of Section 2.2. Finally, if time
is fundamental, the introduction of time uncertainty, Section 3,
can be interpreted differently. To this end, e.g. removing of the
integration from eq.(9) provides the state $\rho(t) \vert
\Psi(t)\rangle\langle\Psi(t)\vert$, which introduces time as a
classical system, $T$, which extends the quantum system $O + A$.
Then the total system $T + O + A$ appears, at least formally, as a
"hybrid system" [46] (and the references therein) that might link
quantum and relativistic theories in a new way [47]. We observe
that the local-time scheme is richer, both conceptually and
interpretationally; as well as being reducible to certain existing
theories and interpretations of quantum theory. To this end, the
local-time scheme points out a new, fresh foundation of quantum
theory, along with some recent approaches [48]  that also, but not
equivalently, perform in the context of the universally valid
Schr\" odinger law. Mathematically elaborate microscopic models of
realistic physical situations, cf. e.g.  [48] (and the references
therein), are highly welcome in this context.

Our conclusions do not directly apply to the weak-interaction
scenarios (e.g. the weak-measurement and some Markovian open
systems dynamics) that require separate considerations. Mutual
relations between the local times remains intact in the present
paper (but see Ref. [5] for a proposal). Finally, interpretation
of eq.(9) in terms of single system of an ensemble of identical
systems in connection to the above described deeper physical
nature of time might provide a fresh look into the long standing
problem of quantum measurement theory. To this end research is in
progress.

\bigskip

{\bf 7. Conclusion}

\bigskip

The local-time scheme of Kitada straightforwardly derives some
basic results of quantum decoherence and measurement theory yet
for the isolated (closed) many-particle system. At the same time,
high quantum coherence is provided for the few-particle systems.
Non-necessity of state collapse (reduction) and environmental
influence, technical simplicity as well as interpretational
ramifications regarding the deeper physical nature of time exhibit
that the scheme is worth a pursuit in foundations and
interpretation of quantum theory and measurement.

\bigskip

{\bf Acknowledgment}

\bigskip

We are deeply indebted to Hitoshi Kitada for patiently explaining
the foundations of the many-body scattering theory and also of his
local-time theory. We benefited much from discussions with Hitoshi
Kitada, C. Jess Riedel, Allen Francom and Stephen P. King.
Comments on an early draft of the present paper we received from
S\" oren Petrat, \' Angel Rivas and Lajos Di\' osi, for which we
express our gratitude.  We acknowledge financial support by the
Ministry of education, science and technology Serbia under the
grant no 171028 and the EU COST Action MP1006.


\bigskip

{\bf Appendix}

\bigskip

The typical scattering situation is described as follows. In the
laboratory reference system, there is a fixed many-particle target
and e.g. a few-particle  projectile directed to the target.
Detected projectile is assumed to be far from the target that is
described by the limit of infinite time, $t \to \infty$.

 The
many-body scattering is truly complex task. It regards all the
possible decompositions of the scattering particles. More
precisely: of interest are  all the possible scattering of
particles that can be composed of the initially introduced
particles.
 And for
every possible such scenario, one should discard the bound states.
Description of this complex picture amounts to the problem known
as the "problem of asymptotic completeness" in quantum many-body
scattering theory. The important work of Enns opened the door for
a solution of this problem.

 Consider an $N$-particle, isolated (closed) system
$\mathcal{S}$ with the Hamiltonian $H$ and the Hilbert state space
$\mathcal{H}$. Denote the individual particles position and
momentum operators by $x_i$ and $p_i$, respectively: $[x_i, p_j] =
\imath \hbar \delta_{ij}, i,j = 1,2,3,...,N$. The total system can
be divided into clusters where a cluster can consist of arbitrary
number of the constituent particles. Let $\mathcal{S}_b =
\{\mathcal{C}_i, i = 1,2,...,k\} $ be the $b$th structure (cluster
decomposition) of the total system $\mathcal{S}$ with the  number
$k$ of clusters. We call "elementary" the structure in which every
particle is one cluster, $\mathcal{S}_e = \{ \{1\} , \{2\}, ...,
\{N\} \}$--the corresponding number of clusters is, of course, $k
= N$.

Consider a structure $\mathcal{S}_b$ with $k$ clusters,
$\mathcal{S}_b = \{\mathcal{C}_1, \mathcal{C}_2, ...,
\mathcal{C}_k\}$, with $N_i$ particles in the $i$th cluster;
$\sum_i N_i = N$. For every cluster introduce the center of mass
and the Jacobi relative positions: $X^b_{CM i}$ and
$x^{C_{bi}}_l$, respectively, where $l = 1,2,3..., N_i-1$. Then
the intracluster variable is defined for the structure
$\mathcal{S}_b$, $x^b = \{x^{C_{b1}}, x^{C_{b2}}, ...,
x^{C_{bk}}\}$. For the set of the clusters' centers of mass, the
Jacobi variables transformation introduce the total system's
center of mass and the intercluster Jacobi relative variable,
$\{x_{b1}, x_{b2},..., x_{bk}\}$. The related conjugate Jacobi
(momentum) variables, $p^l$ and $p_l$, and the commutators
$[x^l_i, p^{l'}_j] = \imath \hbar \delta_{ij} \delta_{ll'}$, and
analogously for the intracluster variables. In the position
representation: $\nabla_l \equiv -\imath \hbar {\partial /\partial
x^l_i}$ is canonically conjugate to the position multiplicative
variable $x^l_i$. In the position representation: $x^l \in
\mathcal{R}^{3(N-k)}$ and $x_l \in \mathcal{R}^{3(k-1)}$. Then the
total Hilbert state space, in the standard functional analysis
notation,
 $\mathcal{H} = L^2(\mathcal{R}^{3N})$, can be factorzied:

 \begin{equation}
\mathcal{H} = \mathcal{H}_{CM} \otimes \mathcal{H}_b \otimes
\mathcal{H}^b,
 \end{equation}

 \noindent which is eq.(3) in the main text.
 By omitting the total $CM$ system from consideration, eq.(56)
 reduces to:

\begin{equation}
\mathcal{H} =  \mathcal{H}_b \otimes \mathcal{H}^b,
 \end{equation}

 \noindent that is eq.(4) in the main text. For $b \neq
 b'$, $\mathcal{H}_b \neq \mathcal{H}_{b'}$ and $\mathcal{H}^b \neq \mathcal{H}^{b'}$, while  $\mathcal{H}_b \otimes \mathcal{H}^{b} = \mathcal{H}_{b'} \otimes \mathcal{H}^{b'}$.

The Hamiltonian for the total system $\mathcal{S}$ and for the
"elementary" structure (with  $x_{ij} = x_i - x_j$):

\begin{equation}
H = \sum_{i=1}^N T_i + \sum_{i\neq j = 1}^N V(\vert x_{ij}\vert),
\end{equation}

\noindent where $T$ stands for the kinetic term, and the
potentials $V$ are the pairwise interactions.

For the $b$th structure, bearing in mind the factorization
eq.(58), the Hamiltonian reads [3,4]:

\begin{equation}
H = T_b \otimes I^b + I_b \otimes H^b_{\circ} + V^{(b)};
\end{equation}

\noindent in eq.(59): $T$ stands for the kinetic term, $H_{\circ}$
 for the "self-Hamiltonian"  and $V^{(b)}$ encapsulates all the interaction
terms for the two factor spaces of the $b$th structure.

Removing the bound states from consideration is managed as
follows. The projector $P_b$ is introduced for the pure point
spectrum of $H^b_{\circ}$. Let us now introduce the "small"
projectors $P_b^M$, $M=1,2,3,...$ such that: (i) $P_b^M P_b =
P_b^M$ and (ii) $s-\lim_{M \to \infty} P_b^M = P_b$. From those
projectors the following operator is built: $\tilde{P}_b^{M_b^m} =
P_b^{M_k} \hat{P}_b^{\hat{M}_b^m}$, where the limit $m \to \pm
\infty$ is equivalent with $M \to \infty$ and the number of
clusters in the $b$th structure is $k$. Without entering the
details, the projector $\hat{P}_b^{\hat{M}_b^m}$ projects onto the
pure continuous spectrum of the Hamiltonian for the structure $b$.
Then Enss was able to prove a theorem that can be concisely
presented by eq.(5). This subtle procedure and the proof can be
found in Refs.[1-4] in the main text..

\end{document}